\documentclass[pra,aps,psfig,epsfig,showpacs,showkeys,floatfix,superscriptaddress]{revtex4-1}
\usepackage{color}

\usepackage{colordvi}

\usepackage{graphicx,amssymb,epsfig,amsmath}
\usepackage{epstopdf}

\begin{document}
\title{Correlation Dynamics of Dipolar Bosons in 1D Triple Well Optical Lattice}
\author{Sangita Bera}
\affiliation{ Department of Physics, Presidency University, 86/1 College Street,
Kolkata 700073, India}
\author{Luca Salasnich}
\affiliation{Department of Physics and Astronomy ``Galileo Galilei'', University of Padova Via Marzolo 8, 35131~Padova,~Italy}
\author{Barnali Chakrabarti}

\affiliation{ Department of Physics, Presidency University, 86/1 College Street,
Kolkata 700073, India}


\begin{abstract}
The concept of spontaneous symmetry breaking and off-diagonal long-range order (ODLRO) are associated with Bose--Einstein condensation. However,   as in the system of reduced dimension   the effect of quantum fluctuation is dominating, the concept of ODLRO becomes more interesting, especially for the long-range interaction. In the present manuscript, we study the correlation dynamics triggered by lattice depth quench in a system of three dipolar bosons in a 1D triple-well optical lattice from the first principle using the multiconfigurational time-dependent Hartree method for bosons (MCTDHB). Our main motivation is to explore how ODLRO develops and decays with time when the system is brought out-of-equilibrium by a sudden change in the lattice depth. We compare results of dipolar bosons with contact interaction. For forward quench $(V_{f} > V_{i})$, the system exhibits the collapse--revival dynamics in the time evolution of normalized first- and second-order Glauber's correlation function, time evolution of Shannon information entropy both for the contact as well as for the dipolar interaction which is reminiscent of the one observed in Greiner's {experiment} [Nature, {415}~(2002)]. 
 We define the collapse and revival time ratio as the figure of merit ($\tau$) which can uniquely distinguish the timescale of dynamics for dipolar interaction from that of contact interaction. In the reverse quench process $(V_{i} > V_{f})$, for dipolar interaction, the dynamics is complex and the system does not exhibit any definite time scale of evolution, whereas the system with contact interaction exhibits collapse--revival dynamics with a definite time-scale. The long-range repulsive tail in the dipolar interaction inhibits the spreading of correlation across the lattice sites.
\end{abstract}

\maketitle

\section{Introduction} \label{intro}
 The non-equilibrium dynamics of isolated quantum systems have gained enormous attention due to recent experiments with cold atoms in optical lattice. One-dimensional Bose gases have proven to be a versatile test bed for the study of quantum many-body physics out of equilibrium. The cold  atoms in optical lattice  offer  precise control over  many system parameters and the high resolution image technique allows   probing  their correlation dynamics~\cite{bakr,hung}. Over the last few years, the forefront research in this direction characterizes many-body systems mainly with contact interaction~\cite{natphys:13,chen,nat440:14,nat415:10,nat419:11,nat465:12,lan,pra99,prb99}. The~onset of thermalization  caused by the interparticle interaction has been widely discussed in many contexts~\cite{miyake,natu85,chen,science341,gring,fischer,lahaye}. The  necessary condition for thermalization is the statistical relaxation. Many-body systems relaxing to equilibrium can exhibit complex dynamics.   In the context of the pioneering experiment of Greiner collapse--revival dynamics between the superfluid $(SF)$ and Mott insulator $(MI)$ phase is observed in the interference pattern~\cite{nat415:10,nat419:11}. \par
Although  many theoretical works exist to characterize the dynamical process, less attention has been paid to probe the effect of long-range interaction in the time scale of dynamical evolution. Out of equilibrium dynamics of quantum systems for long-range interaction has been studied in detail for lattice bosons and spin~\cite{sanchez}. The use of ultracold atoms with dipole--dipole interaction is the most popular tool to understand the physics of long-range~\cite{lahaye,baranov}. Dipolar atoms in quasi-one-dimensional trap are most amenable experimentally and provide rich many-body physics not seen in three dimension~\cite{13_langen,15_langen,16_langen,Zollner_11pra,astrakharchik_08a,astrakharchik_08b,Deuretzbacher_10,arkhipov05,imambekov}. The dipolar atoms in a triple well have already been explored using the Bose--Hubbard model and mean-field theory~\cite{50_budha,51_budha,52_budha,53_budha,54_budha,55_budha,56_budha}. Strongly interacting dipolar bosons have recently been studied by multiconfigurational time-dependent Hartree method for bosons (MCTDHB)~\cite{budha_order} and different quantum phases are addressed. It is also stressed that as the Bose--Hubbard model is unable to address the strongly interacting bosons in shallow lattice, it is indeed necessary to employ a general many-body approach. \par
In our present setup, we consider $N=3$ dipolar bosons in 1D triple-well optical lattice. We initially prepare the system in the $SF$ phase and quench it to the $MI$ phase by sudden increase in the lattice depth. The corresponding dynamics is simulated in the first principle by solving the time-dependent Schr\"odinger equation using MCTDHB method~\cite{mctdhb:14a,mctdhb:14b,PRL99}. It is   a general many-body method capable of addressing our system properly and its implementation in the MCTDH-X package~\cite{axelN1,axelN2,axelN3} has been successfully employed in several earlier works~\cite{ref4,ref5,ref6,PRL99,jpb_peter1,jpb_peter2,ofir76,jpb_peter3,PRA_peter1,PRX_axel,pra2015,roy97,bera18}. We explore the transition from $SF$ to $MI$ phase by analyzing the normalized first- and second-order Glauber's correlation function and dynamics of Shannon information entropy. The observed collapse--revival dynamics is reminiscent of Greiner's experiment~\cite{nat415:10,nat419:11}. We redo the simulation with contact interaction, which also exhibits collapse--revival dynamics but in different time scale. We define figure of merit $\tau$ as the ratio of collapse time and revival time. For both kinds of interaction, $\tau$ does not change with the lattice depth, remaining consistent with the experimental result~\cite{exp_merit}. \par
We also do the reverse quench---the system initially prepared in the $MI$ phase is quenched to $SF$ phase by sudden decrease in lattice depth. We observe contrast between the dipolar and contact interaction in the reverse quench. For contact interaction, the system revives to $SF$ phase, however dipolar bosons exhibit very complicated dynamics. The huge correlation build-up within the individual lattice sites for dipolar interaction is not distributed over the whole lattice---the system does not turn back to $SF$ phase.
\begin{figure}[hpbt]
	\begin{center}
		\includegraphics[height=.75\textwidth, width=0.8\textwidth]{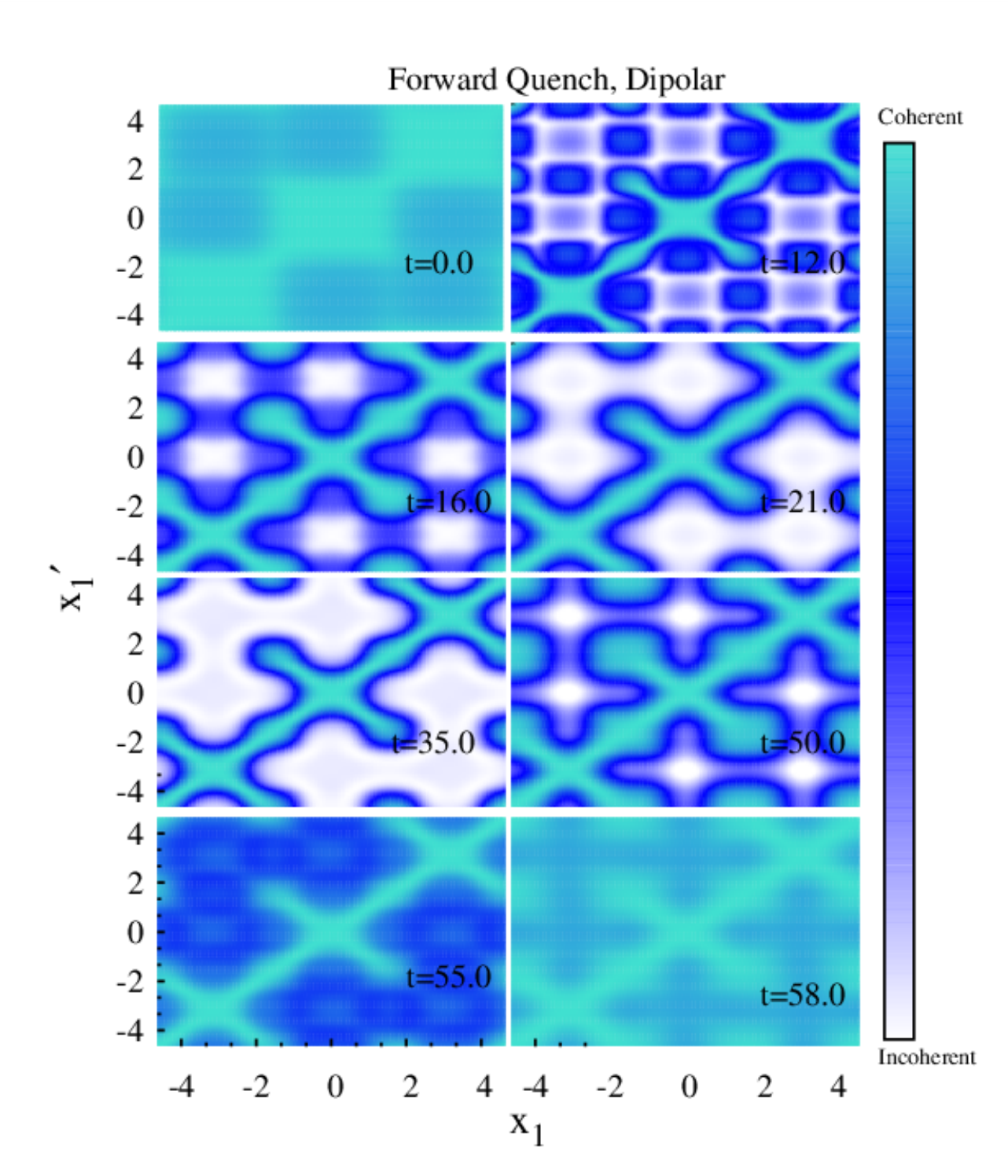}
	\end{center}
	\caption{Time evolution of the normalized first-order Glauber's correlation function $\vert g^{(1)}(x_{1}^{\prime},x_{1};t)\vert^{2}$ for forward lattice depth quench from $V_{i}=3.0$ to $V_{f}=10.0$ for dipolar interaction. We observe collapse ($SF$ $\rightarrow$ $MI$)--revival ($MI$ $\rightarrow$ $SF$) dynamics. See the text for details. All   quantities are dimensionless.}
	\label{for_dip_g1}
\end{figure}
 \begin{figure}[hpbt]
	\begin{center}
		\includegraphics[height=.5\textwidth, width=0.5\textwidth,angle=-90]{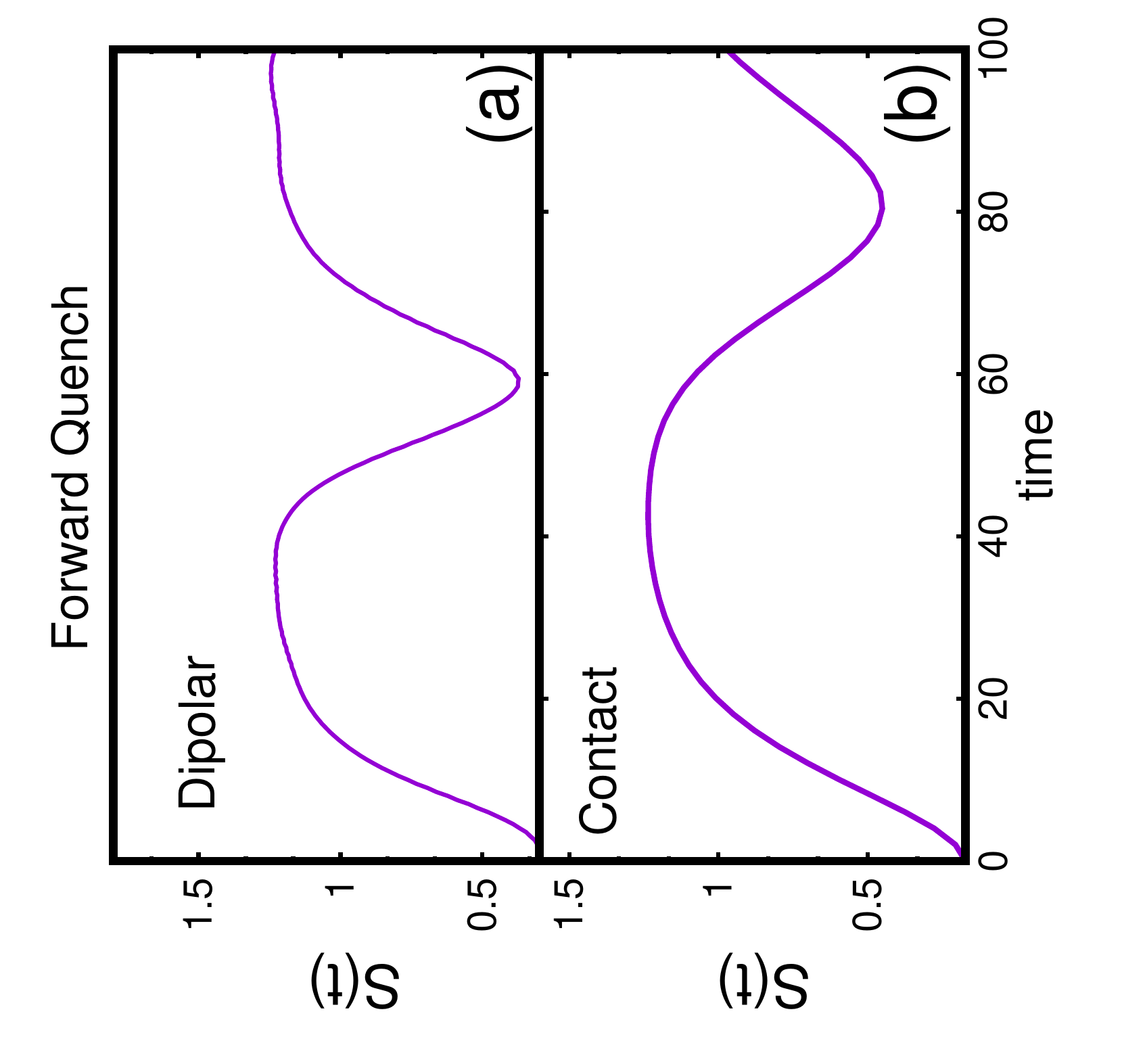}
	\end{center}
	\caption{Time evolution of Shannon information entropy $S(t)$ for forward lattice-depth quench $V_{i}=3.0$ to $V_{f}=10.0$: (\textbf{a})   dipolar interaction; and  (\textbf{b})     contact interaction. In both cases, entropy passes through maximum point ($MI$ phase) and minimum point ($SF$ phase) in their corresponding time scale. All   quantities are dimensionless.}
\label{for_ent}
\end{figure}

\par 
The paper is organized as follows. In Section~\ref{method}, we discuss the Hamiltonian and numerical method to solve the time-dependent Schr\"odinger equation. In Section~\ref{quantity}, we   define the key quantities   we   discuss  in our manuscript. In Section~\ref{result}, we analyze the correlation dynamics for forward and reverse lattice depth quench for dipolar as well as for contact interaction. We draw the conclusion of our paper in Section~\ref{conclusion}. 

\section{Methodology}\label{method}
The evolution of $N$ interacting bosons in 1D is governed by the time-dependent Schr\"odinger equation (TDSE)
\begin{equation}
\hat{H} \psi = i\hbar \frac{\partial \psi}{\partial t}.
\label{TDSE} 
\end{equation}

 The total Hamiltonian is 
\begin{equation}
\hat{H}= \sum_{i}^{N}\hat{h}(x_{i}) + \sum_{i<j=1}^{N}\hat{W}(x_i-x_j),
\label{hamil}
\end{equation}
where $\hat{h}(x_{i}) = -{\hbar^2\over 2m} {d^2\over dx_{i}^2} + V(x_i)$ is the one body Hamiltonian containing the kinetic energy and the external trapping potential. $\hat{W}(x_i-x_j)$ is the interaction potential between two particles at a position $x_{i}$ and $x_{j}$. To solve the TDSE (Equation~(\ref{TDSE})) ,the ansatz for the many-body wave function $\psi(x_1,...,x_N,t) = \langle x_1,...,x_N \vert \psi(t) \rangle$ is taken as a linear combination of time-dependent permanents with time-dependent coefficients
\begin{equation}
\vert \psi(t) \rangle = \sum_{\vec{n}}C_{\vec{n}}(t) \vert \vec{n};t \rangle,
\end{equation}
where, in the second quantized representation,
\begin{equation}
\qquad \vert \vec{n};t\rangle = \prod_{i=1}^M \left[\frac{\left(
	\hat{b}_i^\dagger(t)\right)^{n_i}}{\sqrt{n_i!}} \right] \vert vac \rangle.
\label{second_quan}
\end{equation} 

The summation runs over all possible configurations $N_{conf}= \left(\begin{array}{c} N+M-1 \\ N \end{array}\right)$ and $\{ \vert \vec{n};t \rangle =\vert n_{1},n_{2},..n_{M};t \rangle ;\sum_{i}n_{i}=N \}$. $\hat {b}_{i}^{\dagger}(t)$ creates a boson in the single particle state $\phi_{i}(x;t)$. It is important to emphasize that, in the ansatz, both the expansion coefficient $\lbrace C_{\vec{n}} (t); \sum_i n_i = N \rbrace$ and the orbitals $\lbrace \phi_i(x,t) \rbrace_{i=1}^M$ that build up the permanents are time-dependent fully variationally optimized quantities. \\
\begin{figure}[hpbt]
	\begin{center}
		\includegraphics[height=.65\textwidth, width=0.7\textwidth]{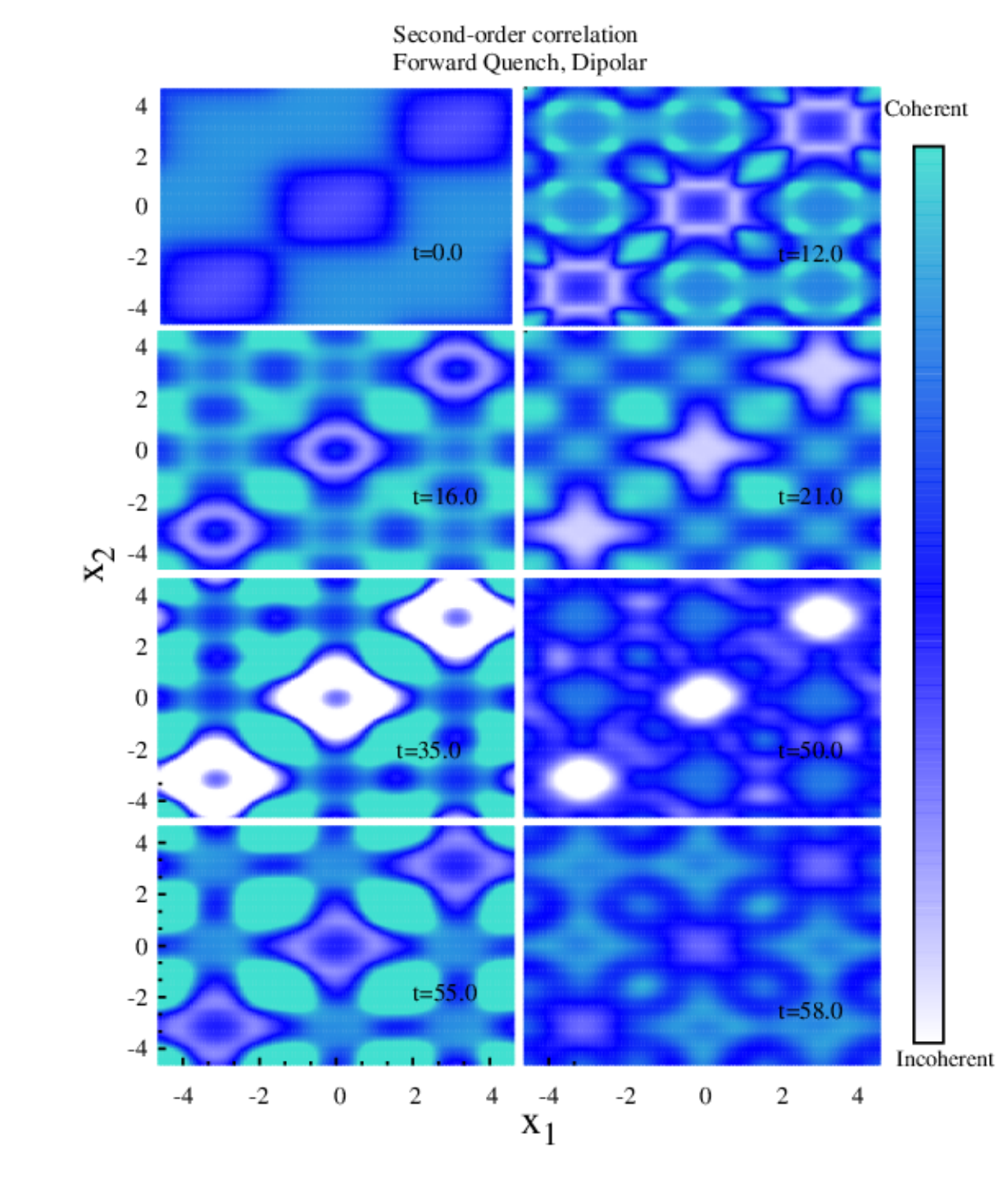}
	\end{center}
	\caption{Time evolution of the normalized second-order Glauber's correlation function $g^{(2)}(x_{1},x_{2};t)$ for forward lattice depth quench from $V_{i}=3.0$ to $V_{f}=10.0$ for dipolar interaction. We observe collapse--revival dynamics (see text). All   quantities are dimensionless.}
	\label{for_dip_g2}
\end{figure}
\unskip
\begin{figure}[hpbt]
	\begin{center}
		\includegraphics[height=.5\textwidth, width=0.5\textwidth,angle=-90]{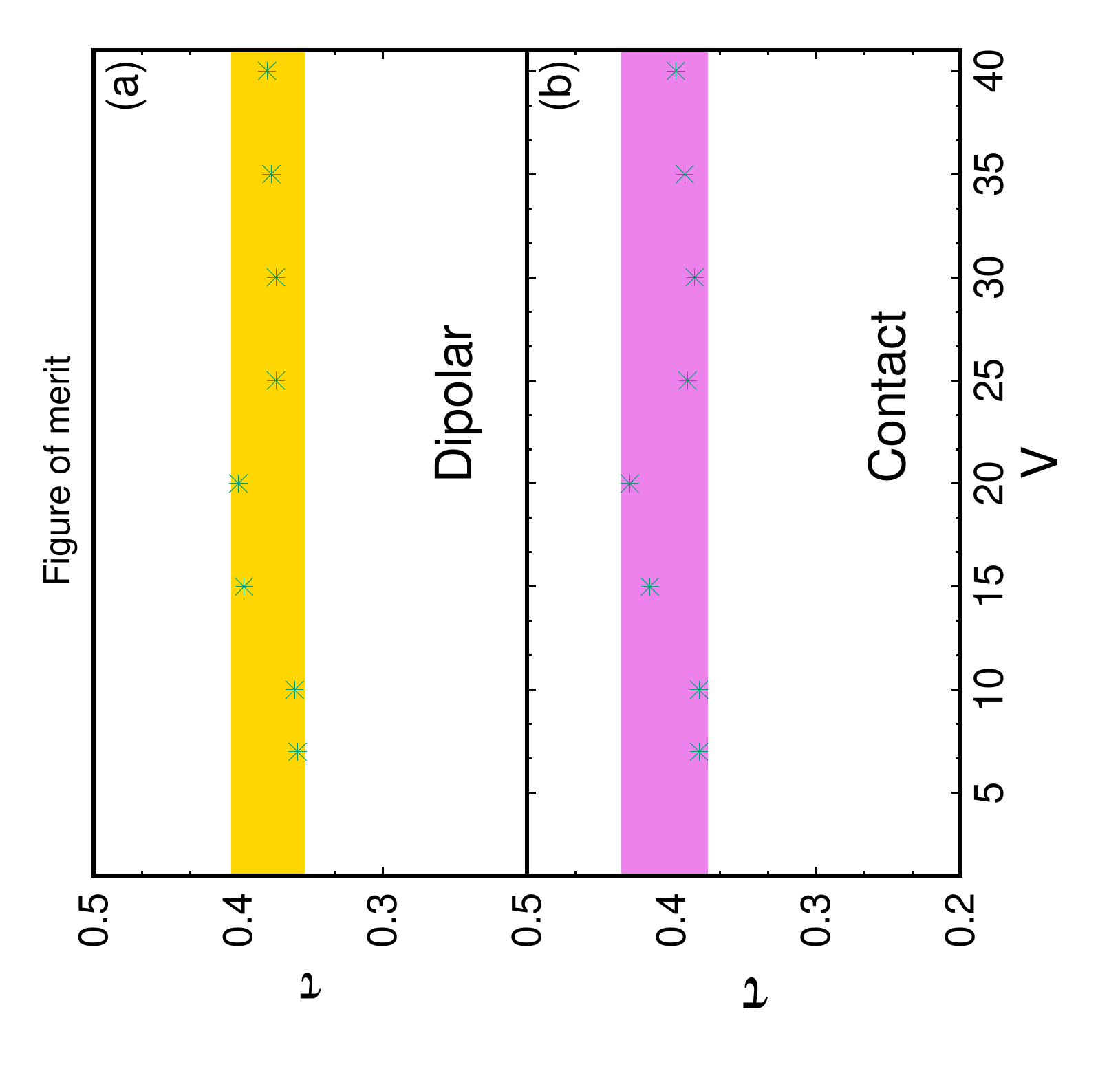}
	\end{center}
	\caption{The ratio of collapse time $t_{collapse}$ to revival time $t_{revival}$ for different lattice depth quench:l (\textbf{a})   for dipolar interaction where the yellow shading has the value $0.379 \pm 0.026$, which is bit higher than the theoretical prediction of $0.128 \pm 0.002$ for Poisson number distribution (Equation (11) of Ref.~\cite{exp_merit}); and (\textbf{b})   for contact interaction, where the purple shading has the width of $0.405 \pm 0.03$, which is bit higher than the prediction given above. In both cases, $\tau$ is independent of lattice depth. }
\label{fig_of_merit}
\end{figure}
 MCTDHB theory is established as the most efficient way to solve the time-dependent many-body problems of interacting bosons accurately and has been applied for a wide set of problems~\cite{ref4,ref5,ref6,PRL99,jpb_peter1,jpb_peter2,ofir76,jpb_peter3,PRA_peter1,PRX_axel,pra2015,roy97,bera18}. In the limit of $M \rightarrow \infty$, the set of permanents $\{ \vert \vec{n};t \rangle \}$ spans the complete $N$ boson Hilbert space. As the permanents are time-dependent, a given degree of accuracy is reached with a shorter expansion as compared to the time-independent basis. To solve the TDSE for the wave function $\vert \psi(x,t) \rangle$, one~needs to determine the evolution of the coefficients $\lbrace C_{\vec{n}} (t); \sum_i n_i = N \rbrace$ and orbitals $\lbrace \phi_i(x,t) \rbrace_{i=1}^M$ with time. Their equations of motion are derived requiring the stationarity of the action functional with respect to the variations of the time-dependent coefficients and the set of time-dependent orbitals. The set of nonlinear integro-differential equations are simultaneously solved by the recursive MCTDHB (R-MCTDHB) package~\cite{axelN1,axelN2,axelN3}. It is stressed that a MCTDHB is much more accurate than exact diagonalization (ED) method. In ED, the time-independent basis is employed which are built up from the eigenvalues of a one-body problem and are not further optimized to take into the account the dynamics and  correlations in the system. Conversely, MCTDHB uses time adaptive many-body basis set and its evolution follows from the time-dependent variational principle. Thus, the error resulting from the truncation of Hilbert space is minimized by the basis at any given instant of time. For the present study,   for both the short- and long-range interactions, we keep $M=6$ orbitals to achieve   convergence.  Convergence is ascertained by systematically increasing the number of orbitals and observing no change in the calculated quantities such as energy and one-body density. Additionally,   convergence is further assured when the occupation of the last orbital is negligible.  In our present computation,  we find that $M = 6$ orbitals are adequate to capture the correct physics.
 \begin{figure}[hpbt]
	\begin{center}
		\includegraphics[height=.45\textwidth, width=0.55\textwidth]{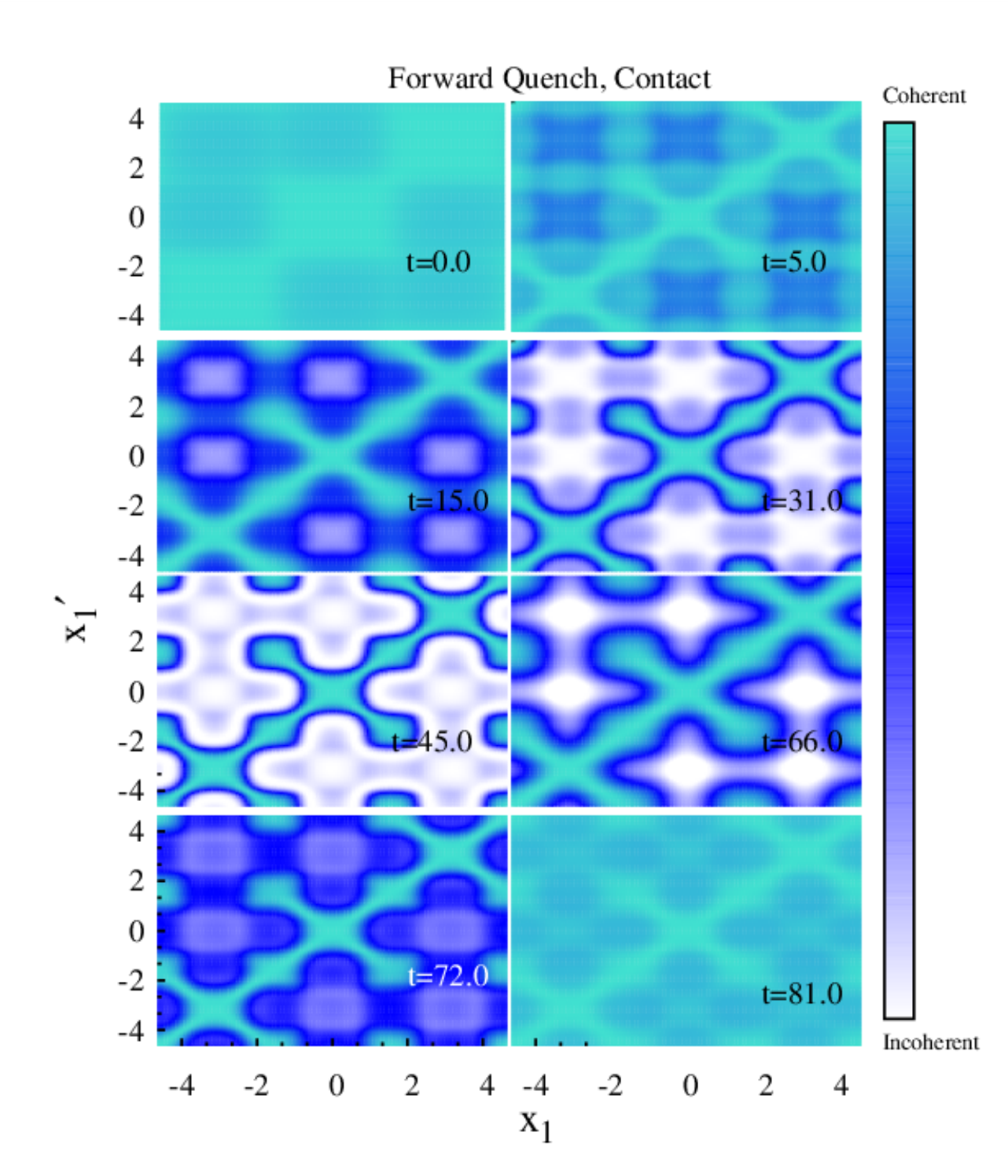}
	\end{center}
	\caption{Time evolution of the normalized first-order Glauber's correlation function $\vert g^{(1)}(x_{1}^{\prime},x_{1};t)\vert^{2}$ for forward lattice depth quench from $V_{i}=3.0$ to $V_{f}=10.0$ for contact interaction. We observe collapse--revival dynamics (see text). All   quantities are dimensionless.}
	\label{for_con_g1}
\end{figure}
\begin{figure}[hpbt]
	\begin{center}
		\includegraphics[height=.45\textwidth, width=0.55\textwidth]{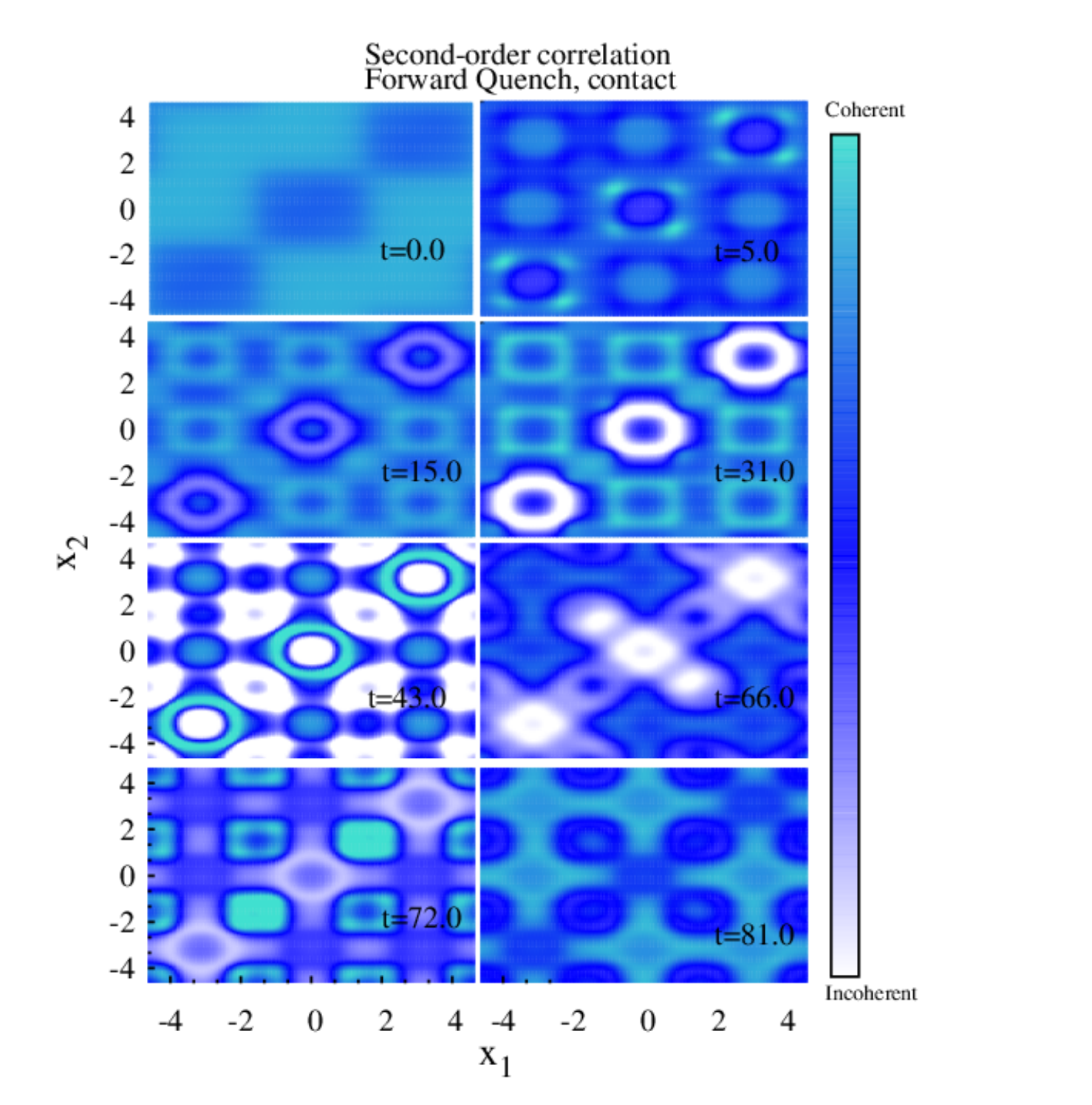}
	\end{center}
	\caption{Time evolution of the normalized second-order Glauber's correlation function $g^{(2)}(x_{1},x_{2};t)$ for forward lattice depth quench from $V_{i}=3.0$ to $V_{f}=10.0$ for contact interaction.  We observe collapse--revival dynamics (see text). All   quantities are dimensionless.}
	\label{for_con_g2}
\end{figure}

\section{Quantities of Interest}\label{quantity} 
The normalized \emph{p}th order correlation function is defined by
\begin{equation}
 g^{(p)}(x_1^{\prime},...,x_p^{\prime},x_1, ...,x_p;t)= 
\frac{\rho^{(p)}(x_1,...,x_p \vert x_1^{\prime},...,x_p^{\prime};t)}
{\sqrt{\prod_{i=1}^p\rho(x_i \vert x_i;t)\rho(x_i^{\prime} \vert x_i^{\prime};t)}},  
\end{equation}
 which defines spatial \emph{p}th order coherence~\cite{36_pra92,37_pra92}, where $\rho^{(p)}(x_1,...,x_p\vert x_1^{\prime},...,x_p^{\prime}; t)$ is the \emph{p}th order reduced density matrix. Although it is possible to define the \emph{p}th order correlation function in the momentum space, in the present manuscript, we report results only for spatial coherence.
The~normalized first-order correlation function $g^{(1)}$ defined as
\begin{equation}
 g^{(1)}(x_1^{\prime},x_1;t)=  \frac{\rho^{(1)}(x_1^{\prime}\vert x_1;t)} {\sqrt{\rho(x_{1}^{\prime};t) \rho(x_{1};t)}},
\label{1st_corr}  
\end{equation} 
which quantifies the degree of first-order coherence. $g^{(1)}<1$ means the visibility of interference fringes in the interference experiment is less than $100\%$ which is referred as loss of coherence.  $g^{(1)}=1$ corresponds to maximal fringe visibility and is referred to as full coherence. The~strength of interaction between the particles affects the correlation; coherence is quickly lost for stronger interparticle interaction.
The corresponding second-order correlation function $g^{(2)}$ is calculated as 
\begin{equation}
 g^{(2)}(x_{1},x_{2};t)= \frac{\rho^{(2)}(x_1,x_2;t)}
{\rho(x_{1};t) \rho(x_{2};t)},
\label{2nd_corr}
\end{equation}
where $\rho^{(2)}$ is the diagonal of the two-body reduced density matrix and $\rho(x,t)$ is the one-body density. $g^{(2)}<1$ is referred to as anti-bunching effect and $g^{(2)}>1$ is termed as bunching effect.   $g^{(2)}=1$ signifies that the measures of two particles at positions $x_{1}$ and $x_{2}$ are stochastically independent. \par
The Shannon information entropy of the one-body density in co-ordinate space is given by 
$ S_x(t) = - \int dx \rho(x,t) ln [\rho(x,t)] $
 and analogously in the conjugate space 
$S_k(t) = - \int dk \rho(k,t) ln [\rho(k,t)]$~\cite{massen,sudip_pra13}.
As the above distributions are related to the one-body density, they are insensitive to correlations that may be present in the many-body state $\vert \psi \rangle$. In MCTDHB theory, as 
$\vert \psi(t) \rangle = \sum_{\vec{n}}C_{\vec{n}}(t) \vert \vec{n};t \rangle$, we define alternative measure of information entropy using the time-dependent coefficients as
 \begin{equation}
S^{info}(t) = -\sum_{\bar{n}}^{}{\vert C_{\bar{n}}(t)\vert}^{2}\ln {\vert C_{\bar{n}}(t)\vert}^{2}.
\label{1BS_c}
\end{equation}

 We term this entropy as coefficient Shannon information entropy (C-SIE). It measures the effective number of basis sets that contribute to a given many-body state. As the mean-field state is a single configurational state only a single coefficient contributes and $S^{info}(t)=0$ for all time $t$,  $S^{info}(t)$ cannot be produced in mean-field theory.

\begin{figure}[hpbt]
	\begin{center}
		\includegraphics[height=.75\textwidth, width=0.8\textwidth]{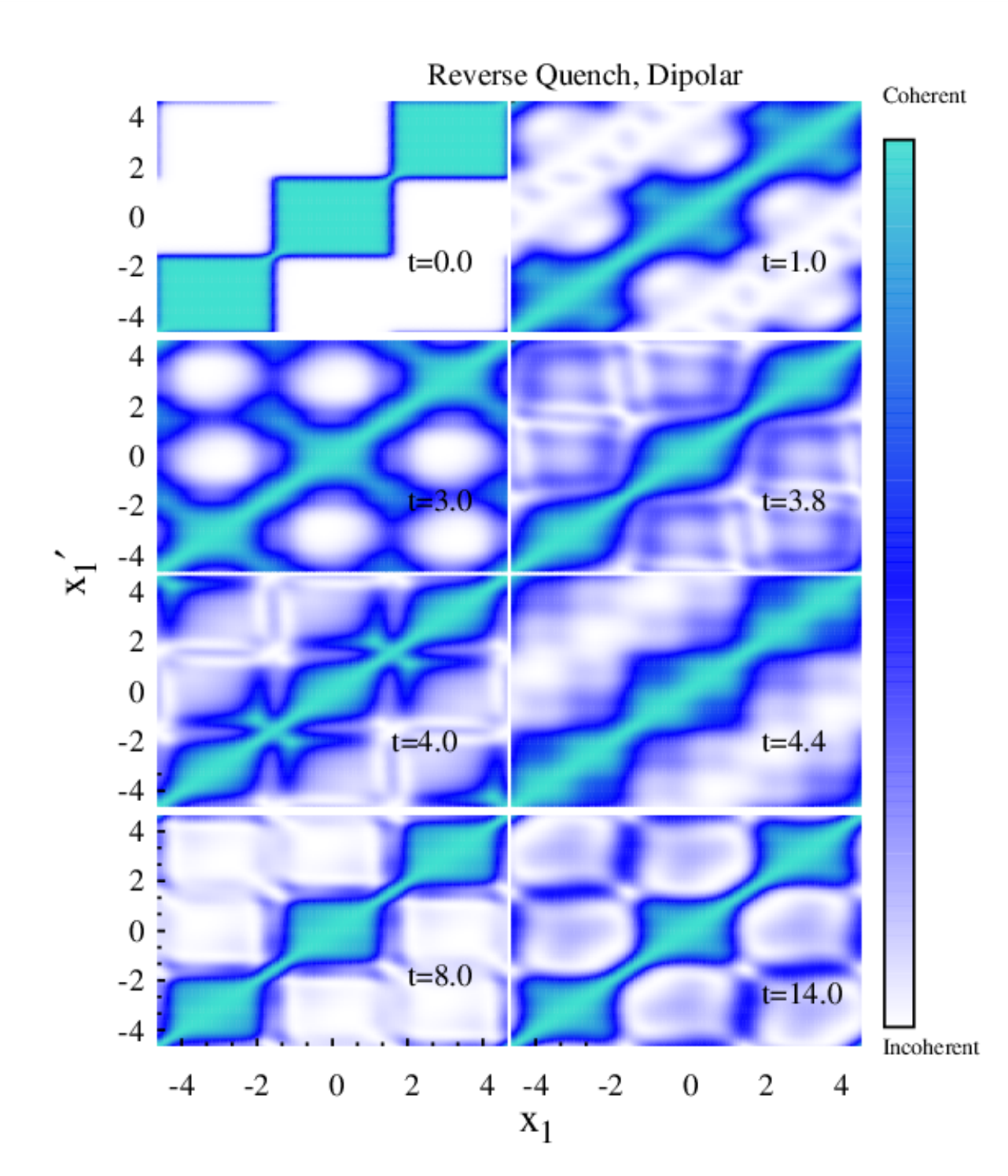}
	\end{center}
	\caption{Time evolution of the normalized first-order Glauber's correlation function $\vert g^{(1)}(x_{1}^{\prime},x_{1};t)\vert^{2}$ for reverse lattice depth quench from $V_{i}=10.0$ to $V_{f}=3.0$ for dipolar interaction. The initial Mott phase builds up some faded off-diagonal correlation, but due to long-range repulsive tail, $SF$ phase is not achieved. All   quantities are dimensionless.}
	\label{rev_dip_g1}
\end{figure}
\begin{figure}[hpbt]
	\begin{center}
		\includegraphics[height=.6\textwidth, width=0.6\textwidth,angle=-90]{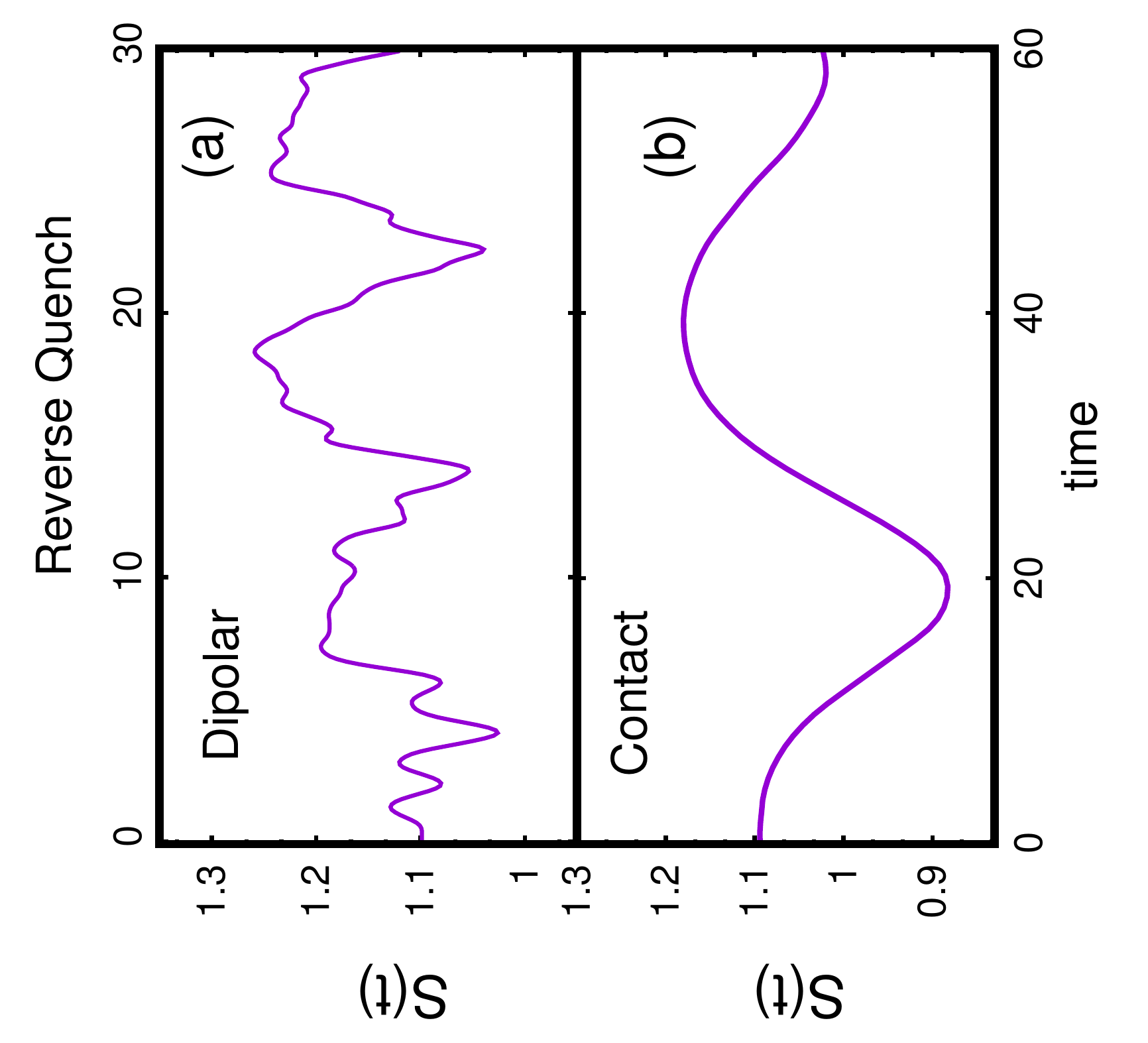}
	\end{center}
	\caption{Time evolution of Shannon information entropy $S(t)$ for reverse lattice-depth quench $V_{i}=10.0$ to $V_{f}=3.0$: (\textbf{a})     dipolar interaction and entropy shows complex dynamics; and (\textbf{b})     contact interaction and the entropy exhibits periodic oscillation. All the quantities are~dimensionless.}
\label{rev_ent}
\end{figure}
\section{Results for Quench Dynamics}\label{result}
The external trapping potential (lattice potential) in the one-body term of the total Hamiltonian in Equation~(\ref{hamil}) can be written as
\begin{equation}
V(x)=Vsin^{2}{(kx)}
\end{equation}
where $V$ is the depth,   $k=\frac{\pi}{d}$ is the lattice wave-vector, and $d$ is the periodicity of the lattice. The~Hamiltonian $H$ is scaled in terms of the recoil energy $E_{R}=\hbar^{2}k^{2}/2m$, $\hbar=m=k=1$, thus~rendering all terms dimensionless. The time is expressed in the units of $\frac{\hbar}{E_{R}}$ and distance is expressed in units of $\frac{1}{k}$. We set our grid to range from $x_{min}=-\frac{3\pi}{2}$ to $x_{max}=\frac{3\pi}{2}$ such that $d=3$ wells are considered. Our setup has $N=3$ bosons in  three wells and filling factor $\nu=1$, which is the elementary building block that exhibits essential dipolar effect of bosons in optical lattice. For the dipolar interaction, 
\begin{equation}
\hat{W}(x_i-x_j) = \frac{g_{d}} {\vert x_i-x_j \vert ^{3} +\alpha_{0}}
\end{equation}
where $g_{d}$ is the dipolar interaction strength and $\alpha_{0}$ is the short-range cut-off to avoid the divergence at $x_{i}=x_{j}$. For contact interaction,
\begin{equation}
\hat{W}(x_{i}-x_{j}) = \lambda \delta (x_{i}-x_{j}),
\end{equation}
where $\lambda$ is the two-body coupling strength between the atoms~\cite{olshanii}. The cut-off parameter $\alpha_{0}$ is chosen by the following normalization
\begin{equation}
\int_{-\infty}^{\infty} \frac{1} {x^3 +\alpha_{0}} dx = \int_{-\infty}^{\infty}\delta(x) dx. 
\end{equation}

For the simulation, we   used the periodic boundary condition and the lower and upper limit of the integration were chosen as   $x_{min}=-4.7$ and $x_{max}=+4.7$, respectively.

\subsection{Forward Quench ($V_{i}=3.0$, $V_{f}=10.0$)}
 We prepared the system in the initial state with the lattice depth $V= 3.0$ and interaction strength $g_{d}=0.01$ which is a $SF$ phase exhibiting both inter- and  intra-well coherence and tunneling is allowed. In the forward quench, we instantaneously increased the lattice depth to $V = 10.0$ keeping $g_{d}$ fixed. The corresponding time evolution of the normalized first-order Glauber's correlation function $\vert g^{(1)}(x_1^{\prime},x_1;t) \vert^{2}$ is plotted in Figure~\ref{for_dip_g1} as a function of two spatial variables $x_{1}^{\prime}$ and $x_{1}$ for various time~$t$. Initially, at time $t=0$, the correlation function remains close to unity for all ($x_{1}^{\prime}$, $x_{1}$), the system is therefore fully coherent. With~increase in time, as fragmentation is built  up in the many-body state, the off-diagonal ($x_{1}^{\prime} \neq x_{1}$) correlation is gradually lost. At time $t=21.0$, when the many-body state is completely fragmented, the correlation function is unity almost exclusively along the diagonal \mbox{($x_{1}^{\prime}$ = $x_{1}$)}. Away from the diagonal ($x_{1}^{\prime} \neq x_{1}$), the correlation function $\vert g^{(1)}(x_1^{\prime},x_1;t) \vert^{2}$ is close to zero. The complete loss of off-diagonal correlation characterizes the fragmented $MI$ state. However,   in long-time dynamics, we observe that the system starts to build up coherence. At time $t=58.0$, both inter- and intra-well coherence is regained ND $\vert g^{(1)}(x_1^{\prime},x_1;t) \vert ^{2}$ becomes unity for all ($x_{1}^{\prime}$, $x_{1}$)---the system revives to $SF$ phase. \par



Since the coherence and correlations are experimentally accessible, it is also instructive to make a connection between the entropy dynamics and the collapse--revival cycle observed in the correlation dynamics. In Figure~\ref{for_ent}a,  we plot the Shannon information entropy (calculated from Equation~(\ref{1BS_c})) as a function of time. We observe a broad maxima from $t =21.0 $ to $t= 54.0 $ , which signifies that the system retains to its maximum entropy state which is a $MI$ phase. $S^{info}(t)$ goes to minimum value at $t=58.0$---the same time when the many-body state revives to $SF$ phase as depicted in Figure~\ref{for_dip_g1}. Thus,  the collapse--revival cycle in coherence dynamics   basically exhibits the maximum--minimum entropy cycle in the evolution of entropy. We define collapse time $t_{collapse}$ in three ways:   
\begin{itemize}

\item[(a)] The system enters completely fragmented $MI$ phase. 
\item[(b)] The Shannon information entropy becomes maximum. 
\item[(c)] The off-diagonal correlation is completely lost.
\end{itemize}
For the present simulation, $t_{collapse}=21.0$---uniquely determined from the above three possible ways. Similarly, revival time $t_{revival}$ can also be calculated in three ways:  
\begin{itemize}
\item[(a)] The system revives to $SF$ Phase. 
\item[(b)] Information entropy reaches its minimum value. 
\item[(c)] The system becomes fully coherent.
\end{itemize}
\vspace{-6pt}

For the present simulation, $t_{revival}=58.0$ was again uniquely determined by the above three criteria.
\begin{figure}[hpbt]
	\begin{center}
		\includegraphics[height=.6\textwidth, width=0.7\textwidth]{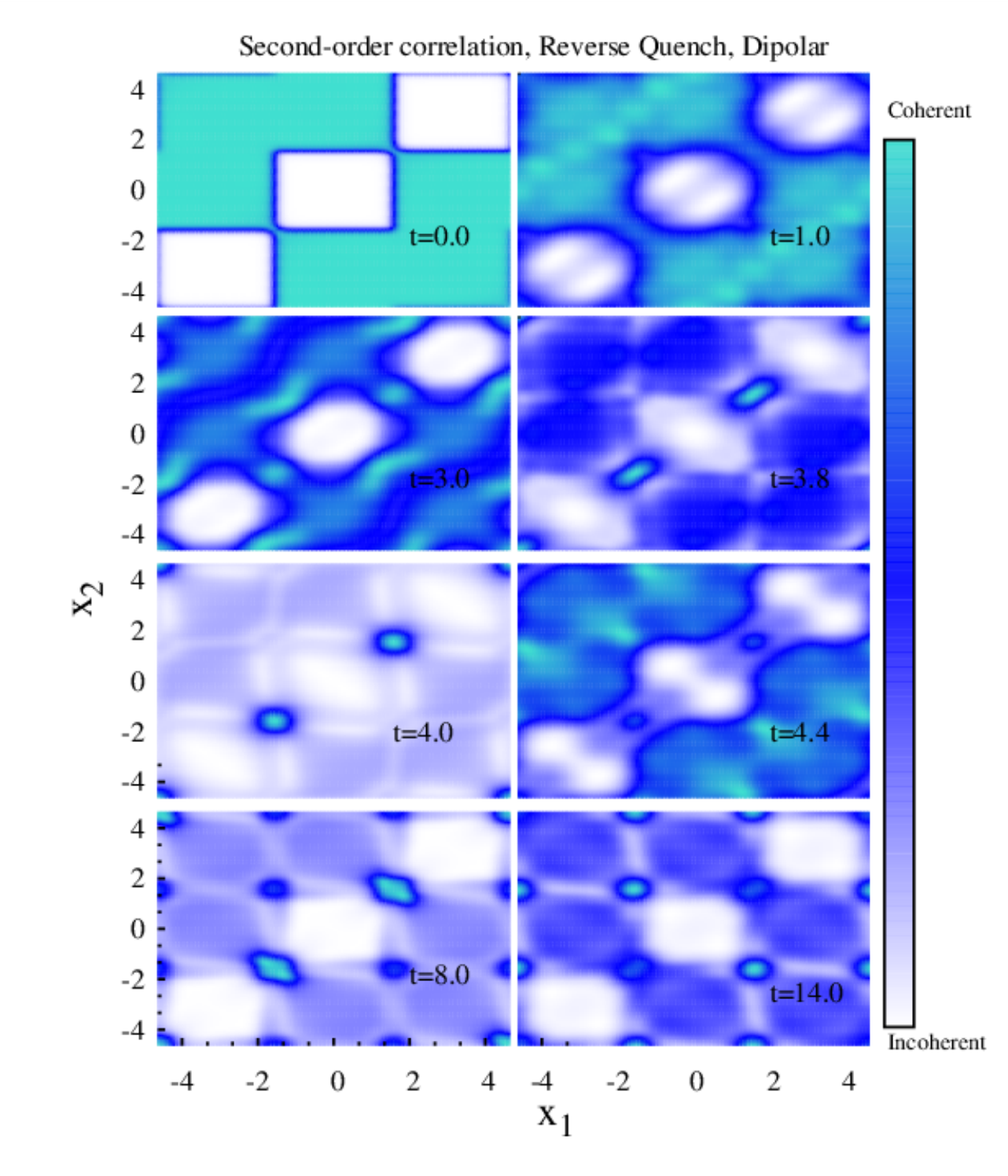}
	\end{center}
	\caption{Time evolution of the normalized second-order Glauber's correlation function $g^{(2)}(x_{1},x_{2};t)$ for reverse lattice depth quench from $V_{i}=10.0$ to $V_{f}=3.0$ for dipolar interaction (see text for details). All   quantities are dimensionless.}
	\label{rev_dip_g2}
\end{figure}
\begin{figure}[hpbt]
	\begin{center}
		\includegraphics[height=.75\textwidth, width=0.8\textwidth]{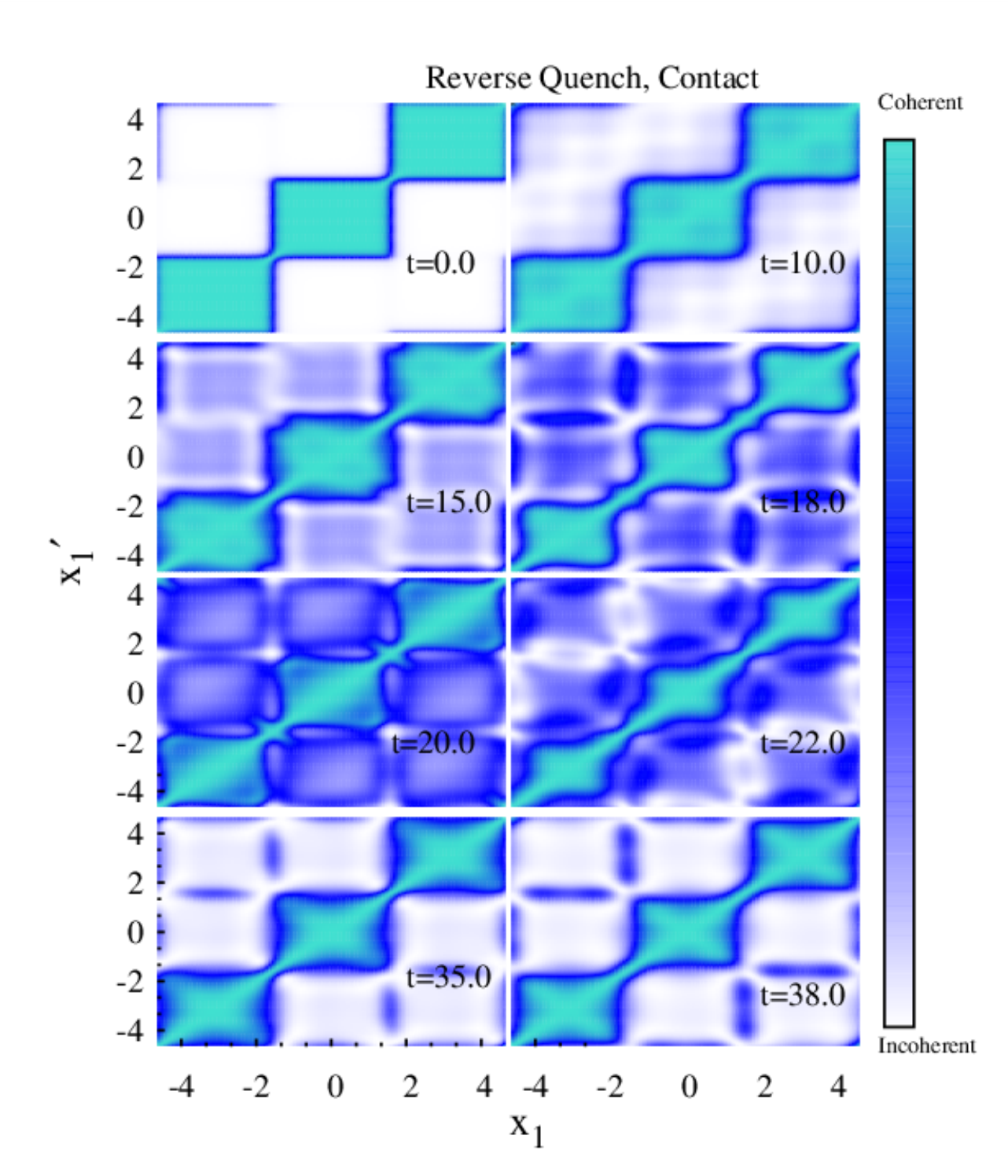}
	\end{center}
	\caption{Time evolution of the normalized first-order Glauber's correlation function $\vert g^{(1)}(x_{1}^{\prime},x_{1};t)\vert^{2}$ for reverse lattice depth quench from $V_{i}=10.0$ to $V_{f}=3.0$ for contact interaction. The initial Mott phase quickly develops off-diagonal correlation throughout the whole lattice and the correlation function for $t=20.0$ depicts $SF$ phase. At $t=38.0$ the system again enters $MI$ phase. All   quantities are dimensionless.}
	\label{rev_con_g1}
\end{figure}
\par The corresponding normalized second-order Glauber's correlation function $g^{(2)}(x_{1},x_{2};t)$ for dipolar interaction is shown in Figure~\ref{for_dip_g2} for different times. Initially, the system maintains second-order coherence for almost all $(x_{1},x_{2})$.  Over time, the diagonal coherence starts to fade out---the probability of detecting two particles along the diagonal decreases with time. Complete deletion of the diagonal is termed as anti-bunching effect---which appears at longer time ($t=35.0)$. However,   at some longer times, the system regains second-order coherence. When we repeated the quench with several larger values of lattice depth, we observed the collapse--revival dynamics is maintained. As $t_{collapse}$ and $t_{revival}$ change in individual quench process, we define the collapse to revival time ratio as a figure of merit ($\tau)$. In~Figure~\ref{fig_of_merit}a, we plot $\tau$ as a function of lattice depth and we observe that $\tau$ is independent of the depth of the lattice potential. This observation is in agreement with previous experimental results (Figure 2 of Ref.~\cite{exp_merit}).

\par
To compare the observed dynamics and its time scale, we repeated the simulation for contact interaction. We observe that s similar collapse--revival dynamics is seen in the first-order coherence (Figure~\ref{for_con_g1}), the system passes from maximum to minimum entropy state (Figure~\ref{for_ent}b) and the anti-bunching effect is also developed in the second-order correlation function (Figure~\ref{for_con_g2}). However~the time scale for contact interaction is much longer. For example, for the same simulation   reported in Figures~\ref{for_dip_g1}--\ref{for_dip_g2}, we observe $t_{collapse}= 31.0$ and $t_{revival}= 81.0$ for contact interaction in Figures~\ref{for_con_g1}--\ref{for_con_g2}. Similar to dipolar interaction, we redid various lattice depth quenches for contact interaction and calculate $\tau$. We plot $\tau$ as a function of lattice depth $V$ in Figure~\ref{fig_of_merit}b, which lies within the same band as depicted for dipolar interaction in Figure~\ref{fig_of_merit}a.
\subsection{Reverse Quench ($V_{i}=10.0$ to $V_{f}=3.0)$}
The dynamics of dipolar bosons becomes more interesting if we do the reverse quench. Now,~we prepare the initial state with lattice depth $V=10.0$ and interaction strength $g_{d}=0.1$, which is a $MI$ phase---strong intra-well coherence is maintained. In this reverse quench, we fix the interaction strength and the lattice depth is instantaneously lowered to $V=3.0$. The corresponding dynamics are presented in Figures~\ref{rev_dip_g1}--\ref{rev_dip_g2}. At $t=0$, the normalized first-order Glauber's correlation function $\vert g^{(1)}(x_1^{\prime},x_1;t) \vert^{2}$ exhibits three separated lobes along the diagonal (Figure~\ref{rev_dip_g1}). Thus, $\vert g^{(1)}(x_1^{\prime},x_1;t) \vert^{2}$ is unity along $x_1^{\prime} = x_1$ and it is zero for all $(x_1^{\prime} \neq x_1)$. It signifies that first-order coherence is maintained within the wells and inter-well coherence is zero---it is a $MI$ state. In the reverse quench process, the~already built up intra-well coherence should be distributed across the lattice as tunneling is allowed. However,~Figure~\ref{rev_dip_g1} shows that first-order coherence exhibits complex dynamics---the~system tries to build up off-diagonal correlation through inter-well tunneling but $SF$ phase is not achieved. When we performed long-time dynamics---the~system passes through complex pictures and $SF$ phase is not regained. The corresponding entropy evolution presented in Figure~\ref{rev_ent}a also shows complex dynamics---the maximum or minimum entropy points are not clear. The corresponding normalized second-order Glauber's correlation function shown in Figure~\ref{rev_dip_g2}  also exhibits that strong anti-bunching effect of the initial state retains with time. Figures~\ref{rev_dip_g1}--\ref{rev_dip_g2} uniquely exhibit the effect of long-range repulsive tail of dipolar interaction which inhibits the spreading of correlation throughout the lattice sites.


\par
\begin{figure}[hpbt]
	\begin{center}
		\includegraphics[height=.75\textwidth, width=0.8\textwidth]{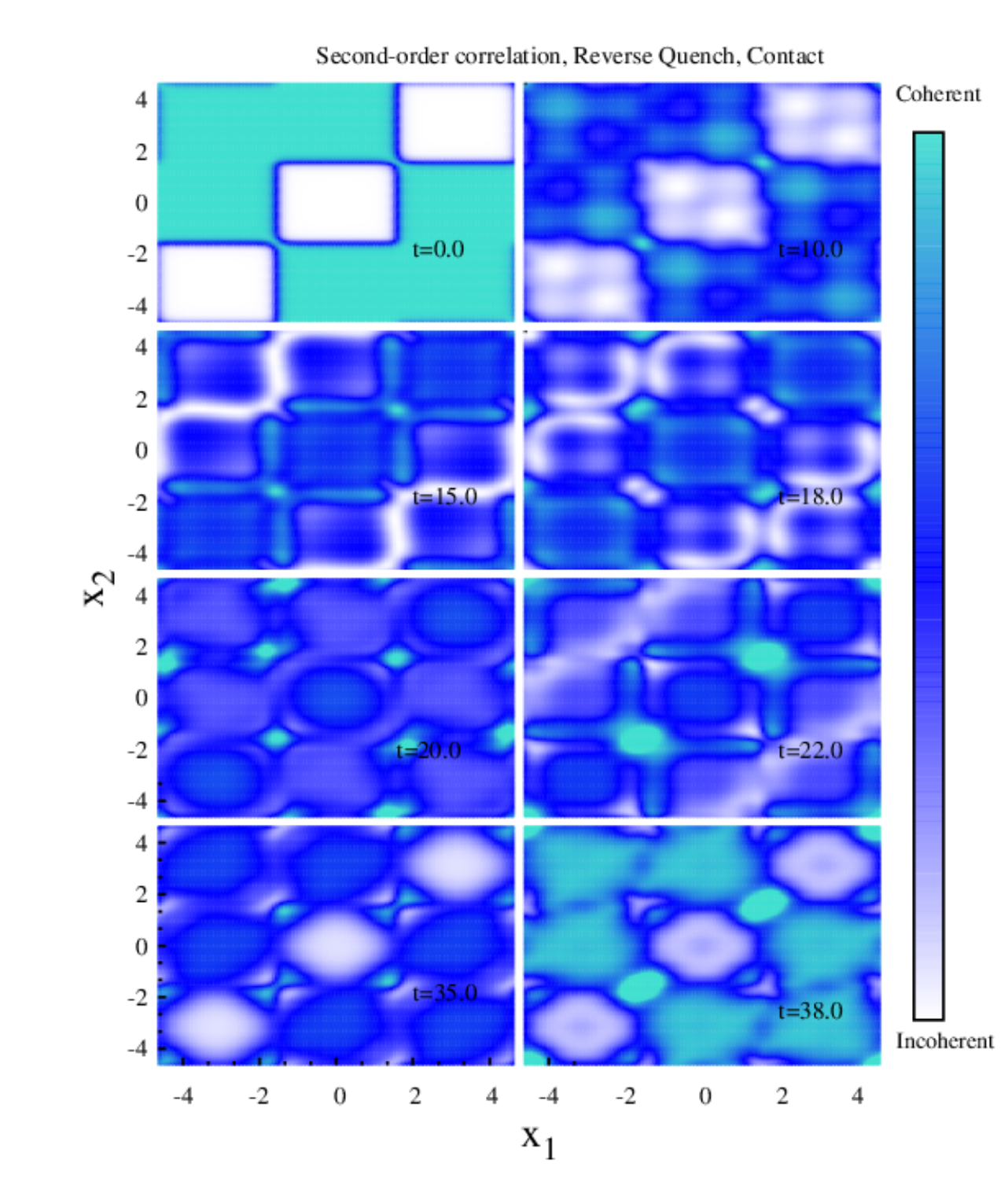}
	\end{center}
	\caption{Time evolution of the normalized second-order Glauber's correlation function $g^{(2)}(x_{1},x_{2};t)$  for reverse lattice depth quench from $V_{i}=10.0$ to $V_{f}=3.0$ for contact interaction (see text for details). All   quantities are dimensionless.}
	\label{rev_con_g2}
\end{figure}

 
It is also interesting to compare the reverse quench dynamics for the contact interaction. The corresponding normalized first-order Glauber's correlation function, entropy dynamics and normalized second-order Glauber's correlation function are presented in Figures~\ref{rev_con_g1},  ~\ref{rev_ent}b and  ~\ref{rev_con_g2}, respectively. All of them exhibit collapse--revival dynamics. In Figure~\ref{rev_con_g1}, we observe that initial Mott phase with exclusive diagonal correlation gradually builds up off-diagonal correlation---it~is a $SF$ phase. At some later time, $MI$ phase comes back. In Figure~\ref{rev_ent}b, we also observe  that, in the revival--collapse cycle---entropy passes through maximum--minimum points. The normalized second-order Glauber's correlation function shown in Figure~\ref{rev_con_g2} also exhibits the revival--collapse scenario. Initially, the system has strong anti-bunching effect across the diagonal, which gradually disappears with time and the system becomes fully second-order coherent. At a later time, the anti-bunching effect develops again.

\par
Thus, our main observations:   (1) The forward quench dynamics both for contact as well as dipolar interaction can be distinguished in their relevant time scale. The long-range repulsive tail of dipolar interaction makes the effective interaction more repulsive,  which corresponds to faster  time scale dynamics for dipolar bosons. (2) However,   we observe contrast behavior in the reverse quench between the contact and dipolar interaction. Although for contact interaction we observe that initial Mott phase at some later time revives to $SF$ phase, for dipolar interaction, the dynamics is very complicated. We are unable to define any time scale of dynamics; the system does not come back to $SF$ phase even in the quite long-time simulation.  
\par
\begin{figure}[hpbt]
	\begin{center}
	\includegraphics[width=14 cm]{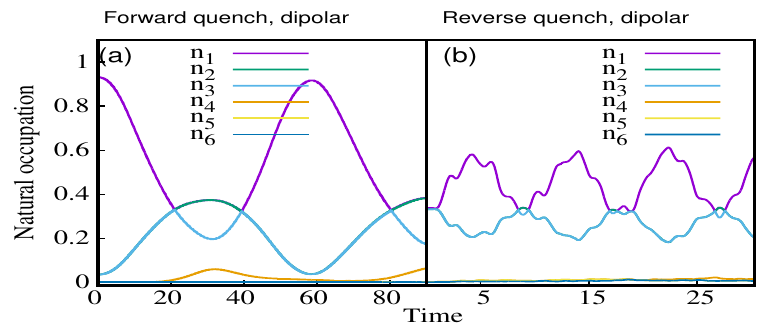}
	\end{center}
	\caption{Time evolution of natural orbitals for dipolar interaction: (\textbf{a})     forward lattice-depth quench $V_{i}=3.0$ to $V_{f}=10.0$; and (\textbf{b})          reverse lattice-depth quench $V_{i}=10.0$ to $V_{f}=3.0$ (see text for details). All   quantities are dimensionless.}
\label{for_rev_occu}
\end{figure}

 The absence of revival in the reverse quench for dipolar interaction can further be explained in terms of population in natural orbitals, as shown in Figure~\ref{for_rev_occu}b. For the forward quench, the initial $SF$ phase is characterized by close to $100 \%$ population in the first orbital (Figure~\ref{for_rev_occu}a). With time, the system fragments: at $t$ = $t_{collapse}$, the system shows three-fold fragmentation, which is a Mott phase; and, at $t$= $t_{revival}$, the first orbital again becomes $100 \%$ populated. Thus at distinct time $SF$ to $MI$ and then $MI$ to $SF$ transition can be identified from the natural orbital occupation. In contrast, for reverse quench process, the system is initially three-fold fragmented; with time the first orbital starts to be more populated with equal decrease in the population in the other two orbitals. However,   the maximum population in the first orbital, as observed in Figure~\ref{for_rev_occu}b, is close to $50 \%$,  thus $SF$ state is not achieved. Thus, the initial Mott phase remains in the Mott phase, and only the diagonal correlation spreads throughout the lattice in a complicated way. However,   as not enough off-diagonal correlation is   built up, $SF$ phase is not revived.  
 
\par Our present simulation is based on the simplest problem of three bosons in three wells, which is the building block of many-body physics. The immediate question in this direction is to investigate the finite size effect keeping the density fixed. We repeated our simulation
for $N = 5$ bosons in five wells keeping filling factor $\nu =1$. We observed identical dynamics as reported for three bosons in three  wells. If we increased the particle number to a sufficiently large value, then   convergence of the measured quantity would be a serious issue. Increasing the particle number requires more and more orbitals to give converged results, which in turn will increase the size of the Hilbert space and numerical simulation will not be possible. 

\section{Conclusions}\label{conclusion}
In this paper, we   study  the quench dynamics of 1D dipolar bosons in a triple-well optical lattice from the first principle general quantum many-body perspective utilizing the MCTDHB method for both  shallow and deep optical lattices. The comparison with the contact interaction is also presented. For forward lattice depth quench for both the contact and dipolar interaction, we observe collapse--revival dynamics in the time evolution of normalized first-order Glauber's correlation function. The observed dynamics is further linked with the production of many-body Shannon information entropy. However,   both the collapse and revival time reported for dipolar interaction are significantly smaller than those corresponding to the contact interaction. We define figure of merit $\tau$ as the ratio of collapse time to revival time, which remains constant for different lattice depth quench. For the reverse quench, the system with contact interaction again passes through several $SF$ or $MI$ phases which are determined as collapse--revival cycle in the entropy dynamics. However,   for dipolar interaction, the initial Mott phase has strong intra-well coherence, which is not distributed across the lattice during its time evolution. Thus, the system exhibits very complex dynamics and the initial Mott phase does not revive $SF$ phase---it is the effect of long-range repulsive tail of the dipolar interaction. 
\par
 The present manuscript mainly focuses on the contrast response of contact and dipolar interaction for reverse lattice depth quench in 1D optical lattice. Although    many published works   report the non-equilibrium quench dynamics for contact interaction, the same for dipolar interaction are few. Recently,  Cevolani {{et al.}}~\cite{sanchez} reported out-of-equilibrium dynamics for long-range interaction. They considered the global interaction quench where both the initial and final states are $SF$. They found that  the  long-range Bose--Hubbard model exhibits the same qualitative behavior as the short-range case. However,   our present manuscript reports both the forward and reverse lattice depth quench where the initial state is $SF$ and $MI$, respectively. Although our observation for the forward quench is qualitatively the same as reported in  \cite{sanchez},   we conclude that reverse quench dynamics for long-range interaction contrast with that for contact interaction, which is not observed yet.






\vspace{6pt}

\acknowledgments{Sangita Bera wants to acknowledge Department of Science and Technology Govt. of India, for the financial support through DST INSPIRE fellowship [2015/IF150245]. Barnali Chakrabarti acknowledges ICTP support where a major portion of the work has been done.}



 \end{document}